\documentclass[11pt,3p,review,authoryear]{elsarticle}

\usepackage{microtype}
\usepackage{graphicx}
\usepackage{subfig}
\usepackage{booktabs}
\usepackage{appendix}
\usepackage{multicol}	
\usepackage{multirow} 
\usepackage{hyperref}
\usepackage{amsmath}
\usepackage{amssymb}
\usepackage{amsthm}
\usepackage{nccmath}
\usepackage{mathtools}
\usepackage{subfig}
\usepackage{adjustbox}
\usepackage{xcolor}

\journal{Machine Learning with Applications}
\begin{document}

\begin{frontmatter}

\title{On the retraining frequency of global models in retail demand forecasting.}

\author[a]{Marco Zanotti\corref{cor}} 
\ead{m.zanotti22@campus.unimib.it}
\address[a]{Department of Economics, Management and Statistics, University of Milano-Bicocca, Milan, Italy}
\cortext[cor]{Corresponding author}

\begin{abstract}
    In an era of increasing computational capabilities and growing environmental consciousness, 
    organizations face a critical challenge in balancing the accuracy of forecasting models with 
    computational efficiency and sustainability. Global forecasting models, lowering the 
    computational time, have gained significant attention over the years. However, the common 
    practice of retraining these models with new observations raises important questions about 
    the costs of forecasting. Using ten different machine learning and deep learning models, we 
    analyzed various retraining scenarios, ranging from continuous updates to no retraining at 
    all, across two large retail demand datasets. We showed that less frequent retraining 
    strategies maintain the forecast accuracy while reducing the computational costs, providing 
    a more sustainable approach to large-scale forecasting. We also found that machine learning
    models are a marginally better choice to reduce the costs of forecasting when coupled with 
    less frequent model retraining strategies as the frequency of the data increases. Our findings 
    challenge the conventional belief that frequent retraining is essential for maintaining 
    forecasting accuracy. Instead, periodic retraining offers a good balance between predictive 
    performance and efficiency, both in the case of point and probabilistic forecasting. These 
    insights provide actionable guidelines for organizations seeking to optimize forecasting 
    pipelines while reducing costs and energy consumption.
\end{abstract}


\end{frontmatter}


\section{Introduction} \label{sec:intro}

Forecasting plays a critical role in decision-making processes across industries, 
from supply chain management to energy demand planning. Traditionally, time series 
forecasting has relied on a Local Modeling approach (LM), meaning that separate models
are trained for each series in isolation. 
However, recent advances have demonstrated the potential of Global forecasting Models 
(GM), a paradigm also known as cross-learning, which in turn consists of fitting a single 
forecasting model on the whole data set \citep{classformeth}. This novel approach allows 
to leverage information across multiple time series to improve accuracy and generalization 
by employing machine learning or deep learning architectures.
Global modeling has affirmed itself as one of the most relevant innovations in time 
series forecasting of recent years \citep{investcrosslearn}, and its success is not only 
due to the demonstrated forecast accuracy in most time series competitions, but also 
to the particular computational benefit it provides. Indeed, while the local approach 
to time series forecasting is potentially beneficial for forecast accuracy, it comes with 
substantial computational overhead. Moreover, since the forecasts are usually produced 
through some cloud computing service that relies on pay-as-you-go pricing 
\citep{occamrazor}, higher computing time and resources directly translate into higher
costs of forecasting for organizations.

The retail industry particularly exemplifies the complexity of modern forecasting 
challenges. Retailers must predict demand for thousands or even millions of products 
across hundreds of stores while accounting for seasonality, promotions, pricing 
strategies, and external factors such as weather and local events. Accurate demand 
forecasts directly impact inventory decisions: overestimation leads to excess inventory 
and waste, particularly critical for perishable goods, while underestimation results in 
stockouts and direct money loss for companies \citep{retailfor}.
Traditional approaches involving separate models for each product-store combination have 
proven increasingly inadequate, especially in contexts where the number of series (SKUs)
to forecast is high, leading to the adoption of global forecasting models that can learn 
patterns across the entire product hierarchy and can drastically reduce the costs of 
producing forecasts.

Nevertheless, despite the growing adoption of global models, it is still common practice 
to update the forecasting models when new observations are available, often motivated by 
the assumption that continuous updates lead to better adaptability to changing patterns 
and more accurate predictions. Unlike local models, global models benefit from 
learning shared dynamics across multiple time series, potentially leading to more 
robust representations that are less sensitive to frequent retraining. 
In the context of local models, the seminal work of \citet{tsupdate1} 
demonstrated that less frequent model updates do not harm forecasting accuracy. 
However, in the context of global models, the effects of retraining are not well 
understood. Is continuous retraining necessary to maintain forecast performance, or 
can global models remain effective without frequent updates? This aspect remains largely 
unexplored in the forecasting literature, and investigating the impact of retraining is 
crucial for both methodological and practical reasons. From a methodological perspective, 
understanding whether global models degrade in performance without frequent retraining 
provides insights into their stability and adaptability. If global models can maintain 
strong predictive accuracy with less frequent updates, it would challenge conventional 
wisdom on the necessity of continuous retraining in forecasting. Moreover, a stable 
forecasting model provides great benefit to the whole decision process the forecasts
are used for. From a practical perspective, frequent retraining has significant 
computational and environmental costs. 
Training large-scale forecasting models requires substantial computational power, 
contributing to energy consumption and carbon emissions \citep{greenai2}.
Indeed, the energy consumption of model training extends beyond direct computational costs. 
In recent years, the environmental impact of machine learning models has become a growing 
concern. Data centers running these models contribute significantly to global carbon 
emissions, with estimates suggesting that training a single large deep learning model can 
emit as much carbon as five cars over their lifetimes \citep{greenai1}. The frequency of 
model retraining multiplies this impact, particularly in large organizations handling 
millions of series, emphasizing the need for more computationally efficient training 
practices. The implication is straightforward: reducing the retraining frequency of 
forecasting models could contribute substantially to sustainability by lowering energy 
consumption, possibly without harming accuracy performance.


\subsection{Research Question}

We aim to extend the results presented in the influential work of \citet{tsupdate1}, 
addressing the question \textit{"Is frequent retraining necessary in the context 
of retail demand global forecasting models?"}. 
Specifically, we study whether skipping retraining when new observations are available 
harms the forecasting performance of global models.
To answer this question, we use the two most recent and comprehensive retail forecasting
datasets, namely the M5 and the VN1 competitions' data.

To generally understand how retraining the forecasting models affects their performance,
we consider ten different global forecasting methods (five from the "classical" machine 
learning domain and five often-used deep neural network architectures), and several possible
retraining scenarios, from continuous retraining to no retraining at all. We also 
explore intermediate periodic retraining strategies to broadly cover the most reasonable
and effective scenarios. 

We also focus on the investigation of trade-offs between accuracy and sustainability, 
in terms of the computational cost of resources to produce the forecasts. This cost is
indeed significant for large-scale applications, like the retail industry, and being
able to reduce (or control) it somehow may result in direct and significant business savings.


\subsection{Contributions}

Our contribution is threefold:

\begin{itemize}

    \item We provide the first comprehensive study of the relationship between retraining 
    frequency and forecast accuracy using 10 different global models, a diverse set of 
    real-world datasets, and focusing on both point and probabilistic forecasting. 
    
    \item We compare different retraining scenarios (e.g., continuous, periodic, and no 
    retraining) on different datasets to quantify the impact of frequent retraining in 
    terms of the cost of forecasting.

    \item We present practical guidelines for organizations and 
    practitioners on when and how often to retrain global forecasting models to balance 
    accuracy and cost.    
    
\end{itemize}

By addressing these points, this paper contributes to both the forecasting and machine 
learning communities, offering insights into the trade-offs between accuracy, efficiency, 
and sustainability in global forecasting models.


\subsection{Overview}

The rest of this paper is organized as follows. 
After a brief review of related works (Section \ref{sec:literature}), 
in Section \ref{sec:exp_des}, we describe the design of the experiment used in our study.
The datasets and their characteristics are presented in \ref{s_sec:data},
the methods adopted for global forecasting are discussed in 
\ref{s_sec:forec_mod}, and the concepts related to model update and
retrain scenario are explained in \ref{s_sec:retrain_scn}.
The evaluation setup through rolling origin validation is presented 
in \ref{s_sec:eval_setup}, and the metrics used to evaluate the model 
performances are shown in \ref{s_sec:perf_met}. 
In Section \ref{sec:results}, we discuss the empirical findings of our
study, including forecast accuracy, computing time, and cost analysis
of the different scenarios.
Finally, Section \ref{sec:conc} contains our summary and conclusions.


\section{Related works} \label{sec:literature}

The literature on cross-learning approach has evolved significantly in recent years. 
Nowadays, most of the works related to time series forecasting include at least some benchmark 
comparison with global models, demonstrating their relevance in the field.
\citet{investcrosslearn} extensively showed their accuracy on the M4 competition dataset, 
\citet{globsim} evaluated the conditions when global forecasting models are competitive, and
\citet{princlocglob} and \citet{princlocglob2} theoretically demonstrated that GM are at least
as accurate as local models with less complexity and without any assumption on the similarity 
of the data.
GM emerged as the most accurate approach in many forecasting areas, such as gas consumption 
\citep{globgas}, electricity demand \citep{globelec}, water demand \citep{globwater}, 
crop production \citep{globcrop}, and retail demand (\citet{globsales1}, \citet{globsales2},
\citet{globsales3}). 
Moreover, GM effectiveness shined during the M5 competition \citep{m5acc}, where tree-based
models leveraging cross-learning were among almost all the top forecasting solutions \citep{fortrees}.
Several techniques like clustering (\citet{globclus1}, \citet{globclus2}), and data 
augmentation \citep{globaug} have been tested to further increase the performance of GM. 
Furthermore, new machine learning \citep{globtree} and deep learning \citep{nbeats} 
architectures specifically designed for cross-learning have been developed.
Finally, a novel area of research is emerging that tries to improve the ability of GM to
capture local patterns \citep{globlocalize}, and their explainability \citep{globexplain}.

From a forecasting evaluation perspective, most of the literature on GM is focused on point 
prediction accuracy, possibly because most machine learning and deep learning methods do 
not directly output probabilistic forecasts \citep{m5unc}. 
Nevertheless, in many forecasting contexts (like supply chain) it is very important to be 
able to produce and evaluate predictions in a 
probabilistic way, being intervals, quantiles, or density based \citep{retailfor}. 

In the context of model retraining and updating strategies, instead, \citet{tsupdate1} is 
the main work related to time series forecasting. The authors explored extensively the 
effects of different retraining scenarios and different forms of model parameter updates on 
the model performance, although they focused on the exponential smoothing family of models 
following the traditional local approach. 
\citet{tsupdate2} briefly discussed retraining in the context of retail demand, but with few 
models and retraining scenarios, and on a proprietary daily dataset only.
Despite the findings being promising, there has been little direct exploration of whether global 
models specifically require frequent retraining or if they can retain competitive accuracy 
with less frequent updates. However, this topic is under consideration of the broader machine 
learning community \citep{greenai4}, advocating for Green AI \citep{greenai2}.

Our study builds upon these existing works, directly investigating the necessity of 
retraining in global forecasting models. By evaluating many different retraining strategies 
and their impact on the forecast accuracy of several global models, we aim to 
provide both theoretical insights and practical recommendations for more sustainable 
forecasting practices.


\section{Experimental design} \label{sec:exp_des} 

This section describes the empirical analysis we conducted to study
whether less frequent retraining scenarios may produce similar accuracy
results compared to the baseline scenario (that with the highest
retraining frequency). First, we describe the datasets used in the
experiments, and then explain the different machine learning and deep
learning models adopted. The performance measures, the several possible 
scenarios, and the strategy used to evaluate the forecasts are also discussed.


\subsection{Datasets} \label{s_sec:data}

For the experiment, we used two retail forecasting datasets: 
the M5 and the VN1 competition datasets. 
The M5 competition was organized by Spyros Makridakis and 
his colleagues as part of the M-competitions series, which 
aimed to compare different forecasting methods in the context
of retail demand \citep{m5comp}. The M5 \citep{m5data} is a well-known and well-studied 
dataset containing 30,490 SKU \(\times\) Store daily time series of Walmart's unit sales of 
products. It covers the sales of three categories of products 
(Food, Hobbies, and Household) sold into ten different stores 
located in three US states (California, Texas, and Wisconsin). 
The time period spans from 2011 to 2016. The time series 
are highly intermittent and are hierarchically organized, 
allowing forecasting at multiple levels, such as individual products, 
product categories, stores, and States. Exogenous information that 
can influence sales, such as product prices, promotions, and 
special events (e.g., holidays) are also available.
The VN1 Forecasting - Accuracy Challenge competition was jointly 
organized by Flieber, Syrup Tech, and SupChains in October 2024, 
and it is the first of its series \citep{vn1data}.
The dataset contains weekly sales of 15,053 products sold from 2020 
until 2024 from e-vendors. In particular, they were mostly online 
from the US, and the products were directly sold to the final consumers.
Contrary to the M5 dataset, where all the products were sold by a single
retailer (Walmart) and from just a few stores, the VN1 dataset includes
product sales of 328 warehouses from 46 different retailers.
As far as we know, we are among the first to test forecasting models 
on this data. These sets of data are the most recent and comprehensive 
time series datasets related to retail demand, allowing for a good
generalization of the results, in particular in the context of demand forecasting.

In both cases, in our experiment, we focused on the most 
disaggregated level (SKU \(\times\) Store), since the potential benefit of 
retraining the models less frequently is much larger at lower levels of aggregation. 
Moreover, for the M5 competition dataset, we did not consider the whole set of 
time series to be able to consistently apply the evaluation setup described
in Sections \ref{s_sec:eval_setup}. 
In particular, for daily data, we kept only time series with at 
least two years of data (730 observations), while for weekly data we 
considered only those that had at least three years of data (157 observations).


\subsection{Forecasting models} \label{s_sec:forec_mod}

In this section, we provide an overview of the global models employed 
for our experiments. 

Let us define \(\mathcal{Y}\) as the set of all available time series in a dataset, 
such that \(Y_i\) represents a single component, and let \(\mathcal{F}\) be 
the set of possible predictive functions, so that \(F\) corresponds to
a single model. 
Without loss of generality, we can assume that all the necessary
information for prediction are contained in \(\mathcal{Y}\).
Under the local approach, predictions for the forecast horizon \(h\) are 
obtained training a model for each time series in the data set, implying
that each time series has its own model, defined by its own parameters values.

\begin{equation}
   Y_i^h = F(Y_i, \theta_i) \mbox{.}
   \label{eq:localmodel}
\end{equation}

On the contrary, following the global modeling framework, forecasts for 
each time series are produced by a model trained on the whole data set.

\begin{equation}
   Y_i^h = F(\mathcal{Y}, \Theta) \mbox{.}
   \label{eq:globalmodel}
\end{equation}

Notice how in the cross-learning methodology, the parameters \(\Theta\) are not 
series-specific, but are common to all time series.

In our experiment, we focused on analyzing the performance of global methods 
only since we are interested in testing whether this approach, 
as opposed to the local one, can benefit from retraining the models
less frequently. Indeed, nowadays, cross-learning is the go-to approach 
for almost all the industries involved with huge time series datasets, 
as in the context of retail demand forecasting, where the forecasts of 
many different SKUs have to be provided regularly \citep{classformeth}. 
For a comprehensive evaluation of global modeling approaches in
demand forecasting, we used both traditional machine learning models 
and cutting-edge deep learning methodologies. The models were selected 
for their established performance in time series forecasting tasks and 
their diverse methodological approaches, enabling a wide comparison and 
avoiding results being dependent on a specific algorithm. 
Furthermore, as benchmarking against established methods is essential in 
time series forecasting, we included two widely recognized global 
forecasting models, Linear (Pooled) Regression and Multi-Layer Perceptron, 
as baselines to evaluate the performance of the proposed approaches.

All the global models were trained using Python under Nixtla's framework \citep{nixtla}. 
The \textit{mlforecast} and the \textit{neuralforecast} libraries were used 
to train machine learning and deep learning models efficiently.

\subsubsection{Machine learning models} \label{ss_sec:ml_mod}

Machine learning models have demonstrated their effectiveness in forecasting
tasks due to their ability to capture non-linear relationships in the data. 
Moreover, they are often easy to train and usually produce
very accurate results, leveraging the cross-learning approach.
In this study, we experimented with Linear (Pooled) Regression and 
four different tree-based methods.

\textit{Linear Regression (LR)} is a classical statistical model 
that assumes a linear relationship between input features and the target variable. 
Despite its simplicity, LR can be effective for time series forecasting
when combined with appropriate feature engineering. The Pooled Regression
is considered a solid benchmark for global model performance evaluation 
\citep{princlocglob, monashrepo}, and it has also proven to
be quite effective \citep{m5accsol2}.

\textit{Random Forest (RF)} is an ensemble learning method based on 
regression trees. It constructs multiple trees during training and aggregates
their predictions through averaging. RF excels in capturing non-linear patterns
and interactions between variables, and it is robust to overfitting, making it 
a strong tree-based model for demand forecasting \citep{randomforest}. It was 
the method used by Amazon until 2015 to forecast e-commerce products demand \citep{fortrees}
\footnote{
    For the cost of computation, the Random Forest model is tested only on the 
    VN1 weekly dataset.
}.

\textit{Extreme Gradient Boosting (XGBoost)} is a gradient-boosted decision tree 
algorithm designed for speed and performance. Iteratively optimizing 
an objective function combines weak learners' predictions to form a solid 
predictive model \citep{xgb}. XGBoost incorporates regularization techniques 
to prevent overfitting, making it particularly effective for datasets with 
complex relationships \citep{xgbtabular}.

\textit{Light Gradient Boosting Machine (LGBM)} is another gradient-boosted 
tree algorithm that emphasizes efficiency and scalability. It uses a 
histogram-based approach to split features and a leaf-wise growth strategy 
to reduce computation time. LGBM is particularly well-suited for large datasets 
and high-dimensional data, often outperforming other boosting algorithms in 
terms of speed and accuracy \citep{lgbm}. In demand forecasting, it is among the
top solutions of all the major competitions (Favorita, Rossmann, M5, and VN1) 
\citep{m5acc, m5unc, m5accsol1, m5uncsol1}.

\textit{Categorical Boosting (CatBoost)} is a gradient-boosted decision tree 
algorithm specifically designed to handle categorical data. By leveraging 
ordered boosting and other innovations, it fosters accuracy and 
enhances generalization \citep{catboost}. CatBoost’s ability to handle categorical features 
without extensive preprocessing makes it advantageous for forecasting 
tasks involving categorical covariates. During the last years, it consistently 
showed top performance on many tabular data studies, becoming the
go-to solution for many practitioners in the field \citep{catbooststudy1, catbooststudy2}.

Machine learning models have the advantage of being easier to train with respect to their deep 
learning alternatives. However, they require extensive and careful feature engineering
to produce accurate results \citep{fortrees}. For this reason, we followed simplified 
versions of the M5 and VN1 top solutions to build time series features.
In particular, we used lags, rolling averages, expanding averages, calendar
features (year, quarter, month, week, day of week, day), and static features
(store, category, location, product identifiers) based on the frequency 
of the dataset. Moreover, for the M5 data, we also used external features 
related to special events as available.
The most relevant hyperparameters of the models were selected based on 
top-performing solutions, otherwise the default values suggested by the 
software providers were adopted. 

\subsubsection{Deep learning models} \label{ss_sec:dl_mod}

Deep learning models have gained prominence in time series forecasting due 
to their capacity to model longer-term dependencies in the data and to 
easily learn hierarchical representations from raw data \citep{deeplearn}. 
In our experiment, we employed five different neural network architectures, 
two well-known methods and three state-of-the-art models.

\textit{Multi-Layer Perceptron (MLP)} is a feedforward neural network 
consisting of an input layer, one or more hidden layers, and an output 
layer. Each layer applies a non-linear activation function to capture 
complex relationships in the data \citep{mlp}.  
MLPs are a versatile and very efficient solution. For this reason, many 
deep learning models specifically created for time series forecasting are MLP-based.
MLP is also considered a solid benchmark for deep learning global 
forecasting models \citep{monashrepo}.

\textit{Recurrent Neural Networks (RNN)} are designed to model sequential 
data by maintaining a hidden state that captures temporal dependencies. 
Standard RNNs, however, suffer from vanishing gradient problems, limiting 
their ability to learn long-term dependencies \citep{rnn}. 
Variants such as Long Short-Term Memory (LSTM) networks and 
Gated Recurrent Units (GRUs) address this limitation and are widely used 
for time series forecasting \citep{lstm}
\footnote{
    We used only the LSTM version, but, for the cost of computation, 
    it was trained only on the VN1 dataset.
}.

For a long time in deep learning, sequence modeling was synonymous with 
recurrent networks, yet several papers have shown that simple convolutional 
architectures can outperform canonical recurrent networks like LSTMs by 
demonstrating longer effective memory. 
\textit{Temporal Convolutional Networks (TCN)} are convolutional architectures 
tailored for sequential data that, by employing causal convolutions and dilations, 
capture long-range dependencies \citep{tcn}.

\textit{Neural Basis Expansion Analysis for Time Series (NBEATS)} is a deep 
learning model specifically designed for time series forecasting. 
It employs a sequence of fully connected layers organized into blocks. Each 
block outputs both trend and seasonal components, enabling the model to 
decompose and predict time series effectively. The network has an interpretable 
configuration that sequentially projects the signal into polynomials 
and harmonic basis to learn trend and seasonality components \citep{nbeats}. 
The \textit{Neural Basis Expansion Analysis with Exogenous (NBEATSx)}, 
allows for the inclusion of exogenous temporal variables available at 
the time of the prediction \citep{nbeatsx}. 
This method has shown state-of-the-art performance on several benchmark datasets
and competitions \citep{m5accsol3}.

\textit{Neural Hierarchical Interpolation for Time Series (NHITS)} builds upon 
the success of NBEATS by incorporating hierarchical interpolation
mechanisms to better capture time hierarchies in time series data. 
Multi-rate input pooling, hierarchical interpolation, and backcast residual 
connections together induce the specialization of the additive predictions in 
different signal bands, reducing memory and computational time, thus 
improving the architecture's parsimony and accuracy \citep{nhits}.

Deep learning models typically do not require extensive feature 
engineering step needed by their machine learning counterpart, since
they create such features (like lags and rolling averages) internally.
Nevertheless, they usually are more difficult to train since they 
have many more hyperparameters to select affecting the forecasting performance \citep{esrnn}. 
We trained the global deep learning models by adding static, calendar, 
and external features only, and relying on the top competitions' solutions to 
set the hyperparameters' values.


\subsection{Retrain scenarios} \label{s_sec:retrain_scn}

To answer our research question, we explored several possible 
retraining scenarios.
A retrain scenario or retrain window, \(r\), is identified as a positive 
integer representing the frequency at which the model is re-trained or updated. 
Essentially, it represents how many data points need to be passed 
before a new training step is performed.
The retrain scenarios strongly depend on the frequency of the 
dataset considered because the frequency drives both the forecast 
horizon and the business review periods. Therefore, we defined two different 
sets of retraining scenarios based on the frequency of the datasets.

\begin{table}[ht]
    \caption{Retraining set, train and test size, frequency, forecast horizon,
    and step size for each dataset. The setup is based on the respective time 
    series frequency.}
    \centering
    \begin{tabular}{lll} 
    \toprule
    & \textbf{M5} & \textbf{VN1} \\
    \midrule
    Frequency (\(f\)) & daily (7) & weekly (52) \\
    Train size (\(n\)) & $\geq 730$ & $\geq 157$ \\
    Test size (\(T\)) & 364 & 52 \\
    Horizon (\(h\)) & 28  & 13 \\
    Step size (\(p\)) & 1 & 1 \\
    Retrain set (\(r\)) & \{7, 14, 21, 30, 60, 90, 120, 150, 180, 364\} & \{1, 2, 3, 4, 6, 8, 10, 13, 26, 52\} \\
    \bottomrule
    \end{tabular}
    \label{tab:retrain_scenario}
\end{table}

For instance, in the case of daily data, \(r = 7\) implies that the 
model is retrained every 7 new observations come in, that is, every week.
Each set contains ten different values, being as exhaustive and computationally 
feasible as possible.

Note that we are training global models, hence the training dataset is
composed by the training set of each time series in the dataset. Since 
the datasets we considered are all aligned, which implies that, at every 
retrain period, we fit the model on a new training data set containing \(r\)
new observations for each time series (where \(r\) is the chosen retrain scenario).
Moreover, note also that we are considering only two different forms of 
model update: a model can either be completely updated when retraining is 
performed, or used as is to produce forecasts for \(r\) subsequent periods.
Therefore, we are not examining the effect of hyperparameter tuning within
each retrain scenario, given the expensive computational cost of this process
and minor changes expected.

The scenario \(r = 1\) is the so-called \textit{continuous} retraining, and it is the 
most expensive since it implies that the model is retrained for every new observation. 
Theoretically, but usually also practically, it should be the most accurate 
forecasting scenario, because the model used to predict has been trained on all 
the available data points up to that moment.
For this reason, we considered this scenario as the benchmark both in terms of
forecasting accuracy and computational cost. 
For daily data only, however, the benchmark scenario is \(r = 7\), because it
is not common to retrain a global model every day, and usually the update is
performed once a week.
The scenario \(r = T\) is the \textit{no retraining} scenario, meaning that the forecasting
model is fitted just once on the initial training set, then it is used to produce
the forecasts for the entire test set \(T\). It is the lowest computationally expensive
but it should also be the lowest accurate. 
All the other scenarios such that \(1 < r < T\) are considered as \textit{periodic} updating
strategies. Both the accuracy and the computing time should be non-increasing functions of \(r\). 


\subsection{Evaluation setup} \label{s_sec:eval_setup}

In time series forecasting, out-of-sample testing is crucial for assessing 
how well a model can generalize to unseen data. This is particularly important 
because patterns in historical data may not fully represent future trends, 
and unexpected changes in the data may occur \citep{tseval}.
As suggested by \citet{tscrossval}, rolling origin evaluation is 
the most widely used and correct method for conducting such tests, offering 
a systematic way to evaluate forecasting models over multiple iterations.
The rolling origin evaluation process begins by dividing the time series into 
two parts: a training set, which is used to build the model, and a test set, 
which is used for evaluation. 
The test set always follows the training set chronologically to maintain the 
temporal order of the data. 
At each step, the model is trained on the training set and generates forecasts 
for a specified number of future points, the forecast horizon \(h\). 
These forecasts are then compared to the actual values in the test set using a 
chosen evaluation metric (see Section \ref{s_sec:perf_met}).
What makes rolling origin evaluation unique is its iterative nature. After each 
forecasting step, the forecast origin (the last point in the training set) is 
shifted forward by a fixed number of periods, the step size. The model is then retrained using 
the updated training set, which may either expand to include all available data 
up to the new forecast origin or remain fixed to a specific window size. 
This process is repeated until the test set is fully utilized, and the overall 
performance of the model is calculated by averaging the results across all the 
iterations.

Compared to the fixed origin evaluation setup, which allows for a single evaluation
step, the rolling origin evaluation is preferable for its ability to provide a more 
comprehensive assessment of a model's performance under different conditions, 
such as varying seasonal patterns, level shifts, or data trends. By simulating multiple 
forecast cycles, indeed, this method reduces the risk of bias that can arise from 
evaluating a model a single time and at a single forecast origin \citep{tscrossval}. 
This approach is beneficial in industries like retail and supply chain 
management, where forecasts need to be continuously updated to reflect real-time 
changes. By simulating multiple scenarios and evaluating a model’s performance 
across different cycles, rolling origin evaluation provides valuable insights into 
the robustness and reliability of the forecasting methods adopted.

In practice, the rolling origin evaluation setup can be adjusted based on the 
data and forecasting goals. For instance, the training set can be fixed to a rolling 
window of the most \(n\) recent observations (fixed window), or it can expand to 
include the full historical data (expanding window) \citep{tscrossval}. 
In real-life applications, most forecast practitioners use this second option, 
especially when the length of the time series is small \citep{fortheopract}. 
In our study, we adopted the expanding window approach because we are interested in 
simulating a forecasting experiment as close as possible to the business reality. 
Moreover, it is also the only reasonable option when dealing with short time series, 
like the case for the weekly VN1 datasets.

The size of the test set \(T\), the length of the forecast horizon \(h\), 
and the step size \(p\), for shifting the forecast origin, are also customizable 
parameters that depend on the specific use case. 
Table \ref{tab:retrain_scenario} shows the parameters selected in
our experiments. Following \citet{tsupdate1}, we set the size of the test set to 
cover at least one complete year so that the evaluation of the different scenarios 
does not depend on intra-year variations, such as any specific season or period of the year.
We selected the forecast horizons based on the type of business decisions that 
usually daily and weekly forecasts support, i.e. operational vs strategic
planning. Finally, the step size was always set equal to one to maximize 
the number of evaluations of each scenario.


\subsection{Performance metrics} \label{s_sec:perf_met}

The accuracy evaluation of point predictive models is a controversial 
topic in time series forecasting: many metrics are available 
to capture models' performances, but no consensus has been reached in the
literature on whether one metric is better than others \citep{forevalds}. 
However, since we are dealing with demand forecasting at SKU-level, 
implying that data intermittency is very much likely, all the metrics
based on absolute or percentage errors are not optimal since they optimize
for the median \citep{evalbestacc}.
Moreover, being the scale possibly different among the series, a scaled 
accuracy measure is preferable.
For this reason, we considered the Root Mean Squared Scaled Error (RMSSE),
proposed by \citet{evalmeasacc}, to evaluate the point forecast accuracy 
of the models. The RMSSE is defined as:

\begin{equation}
   \text{RMSSE} = \sqrt{\frac{\frac{1}{h} \sum_{t=n+1}^{n+h} 
   (y_t - \hat{y}_t)^2}{\frac{1}{n-s} \sum_{t=s+1}^{n} 
   (y_t - y_{t-s})^2}} \mbox{.}
   \label{eq:rmsse}
\end{equation}

The RMSSE measures the relative prediction 
accuracy of a forecasting method by comparing the mean squared errors of 
the prediction and the observed value against the mean squared errors of 
the seasonal naive model.
This metric was the official metric used to assess the performance 
of models in the M5 competition (with \(s = 1\), hence using the naive
model as a benchmark) \citep{m5comp}. 
Lower RMSSE values suggest higher model accuracy.

Our work is not limited to the evaluation of point forecasting
accuracy, but we also aim to assess the probabilistic performance
of the models throughout each different retrain scenario.
We used a scaled version of the Multi-Quantile Loss to comprehensively 
measure the goodness of the probabilistic predictions. The scaled 
Quantile Loss, also known as (scaled) Pinball Loss, and the scaled 
Multi-Quantile Loss are defined as:

\begin{equation}
    \text{SQL} = \frac{1}{h} \frac{\sum_{t = n+1}^{n+h} 
    \left( q \cdot (y_t - \hat{y}_t) \cdot \mathbb{I}_{y_t \geq \hat{y}_t} 
    + (1 - q) \cdot (\hat{y}_t - y_t) \cdot \mathbb{I}_{y_t < \hat{y}_t} 
    \right)}{\frac{1}{n-s} \sum_{t=s+1}^{n} |y_t - y_{t-s}|} \mbox{,}
    \label{eq:sql}
\end{equation}

\begin{equation}
    \text{SMQL} = 
    \frac{1}{|\mathcal{Q}|} \sum_{q \in \mathcal{Q}} \text{SQL}(q) \mbox{,}
    \label{eq:smql}
\end{equation}

where \(\mathcal{Q}\) is the set of possible quantiles.
The QL, being a proper scoring rule, allows for the correct evaluation
of the probabilistic forecasts \citep{evalcountdata}.
The scaling factor is the in-sample, one-step-ahead mean absolute 
error of the seasonal Naive model. The scaled MQL (weighted and with \(s=1\)) 
was also the official metric used to evaluate the overall performance of the 
forecasting methods during the M5 Uncertainty competition \citep{m5unc}.
For daily data we used \(s = 7\), because \(s = 1\) is arguably an irrelevant 
benchmark in daily retail data, which typically exhibits strong day-of-week 
seasonality \citep{retailfor}. Instead, we set \(s = 1\) for weekly data.

Note that we considered a total of 14 different quantiles, \(q\), 
of probability levels \{0.005, 0.025, 0.050, 0.100, 0.150, 0.200,
0.250,
0.750, 0.800, 0.850, 0.900, 0.950, 0.975, 0.995\}. 
Therefore, we studied the 50\% 
60\%, 70\%, 80\%, 90\%, 95\%, and 99\% symmetric prediction intervals.
The 50\% and 60\% symmetric prediction intervals provide a good description of 
the center of the forecast distribution, while the 90\%, 95\%, and 99\% give more 
information about its tails, which in retail demand forecasting problems are 
fundamental to determine the appropriate safety stock levels \citep{evalinventory}. 
These quantiles provide sufficient information about the uncertainty
of the forecasts and allow for the effective description of the whole distribution. 

Finally, we addressed the problem of evaluating the cost associated to 
producing the forecasts by calculating the Computing Time (CT). 
CT is defined as the time in seconds required for training and predicting 
the next \(h\) time steps ahead with the model.
Among the possible measures of complexity, such as the number of parameters, 
number of iterations, model's depth, etc, CT is the most directly related
to the monetary cost of forecasting since the forecasts are usually produced 
through some sort of cloud computing service that relies on pay-as-you-go
fares \citep{tsupdate1}. Hence, CT is analyzed for each forecasting model and 
retraining scenario combination. Lower CT values imply lower forecasting costs.

Note that we related each evaluation metric result to the baseline
retraining scenario, depending on the frequency of the dataset, to be able to
easily compare the performance among different models and retrain windows.
Furthermore, to statistically verify our research question, we tested the 
scenarios' results using the Friedman-Nemenyi test for multiple comparisons 
\citep{testfriednem}.

For the experiments, we used a cloud computing machine NC6s\_v3 hosted on 
Microsoft Azure, with Linux Ubuntu 24 operating system, 1 Graphical Processing 
Unit, 6 Computing Processing Units, 112GB of memory. Parallelization and 
GPUs were used whenever possible within the model libraries.
The \textit{utilsforecast} library was used to evaluate the model performance.


\section{Results and discussion} \label{sec:results}

In general, the models tested performed better, in absolute terms, on the
M5 dataset compared to the VN1 (see the RMSSE and SMQL tables in the 
Supplementary materials).
This may be due to different factors like the dataset size, the frequency 
of the time series, the availability of external regressors to account for 
promotions, events, etc, and the absence of reference hyperparameter values 
to use during the training process.

\begin{figure}
    \centering
    \includegraphics[scale=0.55]{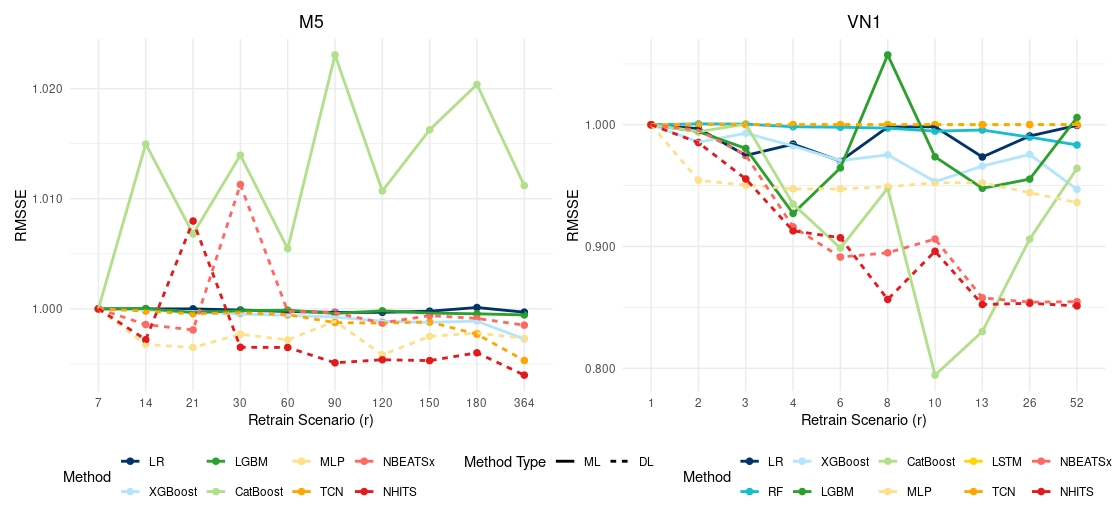}
    \caption{RMSSE results for each method and 
    retrain scenario combination in relative terms 
    with respect to the baseline scenario, \(r = 7\) 
    for the M5 dataset and \(r = 1\) for the VN1 dataset.}
    \label{fig:rmsse_plot}
\end{figure}

Figure \ref{fig:rmsse_plot} shows the forecast accuracy of each model
along the different retraining scenarios for the M5 and the VN1 datasets.
To simplify comparisons, for each dataset, we report the results in relative 
terms with respect to the corresponding baseline scenario, that is
\(r = 7\) for M5 and \(r = 1\) for VN1.
With the exception of the CatBoost model, the RMSSE profiles are very
stable over the retraining frequencies. Indeed, the accuracy remains 
practically the same in the M5 dataset and even improves in the VN1 dataset, 
regardless of the retraining scenario considered. Especially for low 
periodic retraining scenarios, the performance of most global models is 
indistinguishable from the baseline, and even if some deterioration is 
present for higher retraining frequencies, this is less than 5\% also for 
the no retraining setup. These results imply that less frequent retraining 
does not harm the point forecast accuracy of global models.
This can be explained by the fact that, if the demand patterns remain stable without 
significant trends or concept drifts, as is the case for both the M5 and VN1 
datasets, the forecasts will accurately track demand over extended periods.

\begin{figure}
    \centering
    \includegraphics[scale=0.55]{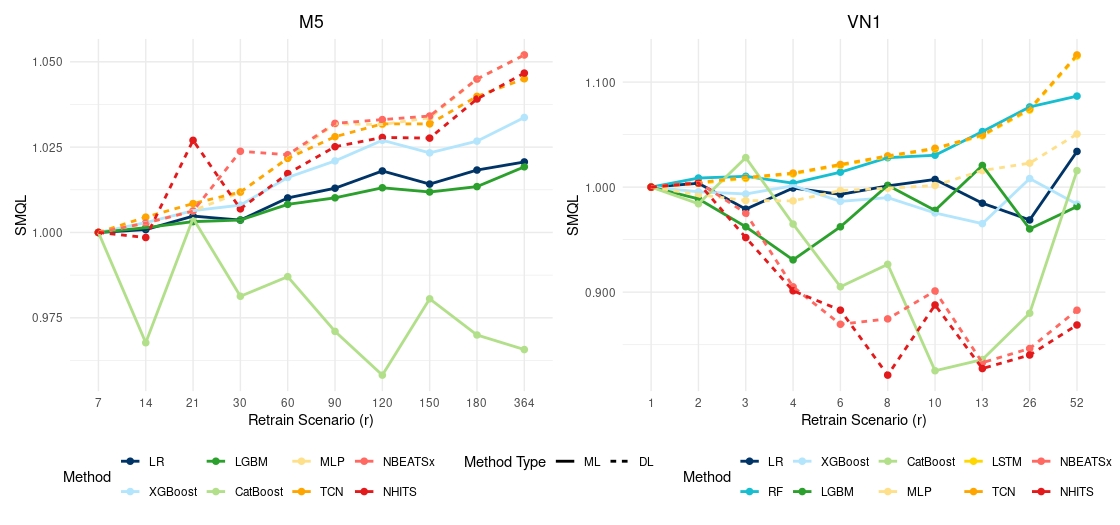}
    \caption{SMQL results for each method and 
    retrain scenario combination in relative terms 
    with respect to the baseline scenario, \(r = 7\) 
    for the M5 dataset and \(r = 1\) for the VN1 dataset.}
    \label{fig:mql_plot}
\end{figure}

In a similar fashion to Figure \ref{fig:rmsse_plot}, Figure \ref{fig:mql_plot} 
summarizes the relative accuracy in a probabilistic forecasting setting, as 
defined in \ref{s_sec:perf_met}. In this context, we observe that, for the M5 
dataset, the accuracy (as measured by Multi Quantile Loss) is clearly an 
increasing function of the retraining scenario, meaning that less frequent 
updates harm the probabilistic forecasting performance of the models. This is 
true regardless of the method used. Nonetheless, the differences in accuracy 
are irrelevant for low retraining scenarios and slightly more pronounced for 
higher retraining levels, but in any case, less than 5-6 percentage points. We also note
a small difference in the performances of machine learning and deep learning 
models, where the formers perform consistently better as the retraining scenario
increases.
For the VN1 dataset, instead, we observe an almost convex relationship between
the accuracy and the retraining period. On average, models' performance improves
for low retraining scenarios and then starts deteriorating around \(r = 4\). 
The only exceptions are NBEATSx and NHITS models, designed also for long-term 
forecasting, which are consistently better compared to the others for longer 
retraining scenarios, both for point and probabilistic forecasting.  

\begin{figure}
    \centering
    \includegraphics[scale=0.55]{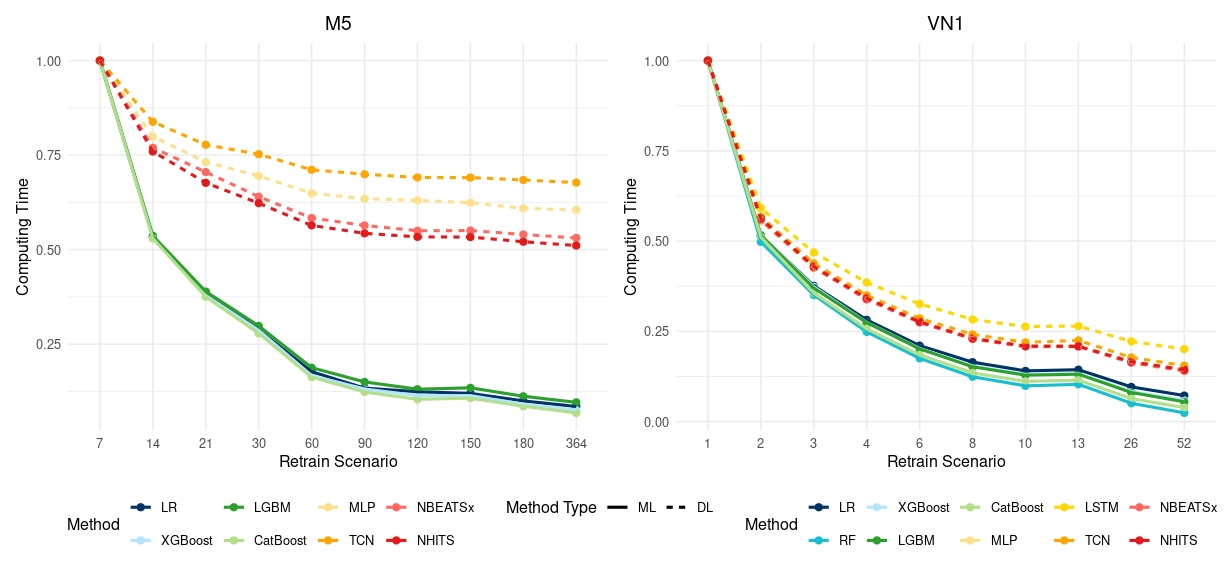}
    \caption{CT results for each method and 
    retrain scenario combination in relative terms 
    with respect to the baseline scenario, \(r = 7\) 
    for the M5 dataset and \(r = 1\) for the VN1 dataset.}
    \label{fig:ct_plot}
\end{figure}

Figure \ref{fig:ct_plot} shows the relative computational time of each 
retraining scenario for the two datasets. We observe that CT decreases 
exponentially as the retraining scenario increases. On average, reductions are 
similar for both, the M5 and the VN1: going from the baseline to the first 
periodic retraining scenario (\(r = 14\) or \(r = 2\)) almost halves CT, 
retraining the models every month often reduces computing time by 75\%, and 
this reduction reaches 90\% in the no retraining scenario, where the models 
are trained just once. 
However, we observe that there is a strong difference between machine 
learning and deep learning models in the M5 data. While the former
continuously undergoes CT reductions, the latter seems to plateau at 
50\%, experiencing much lower gains over increasing retraining scenarios.
As expected, this difference is less evident in the VN1 dataset, since it is
smaller and of lower frequency compared to the M5. 
These results will have direct implications in terms of the costs of forecasting, 
which in turn may guide the choice of the model to adopt.

\begin{figure}
    \centering
    \includegraphics[scale=0.55]{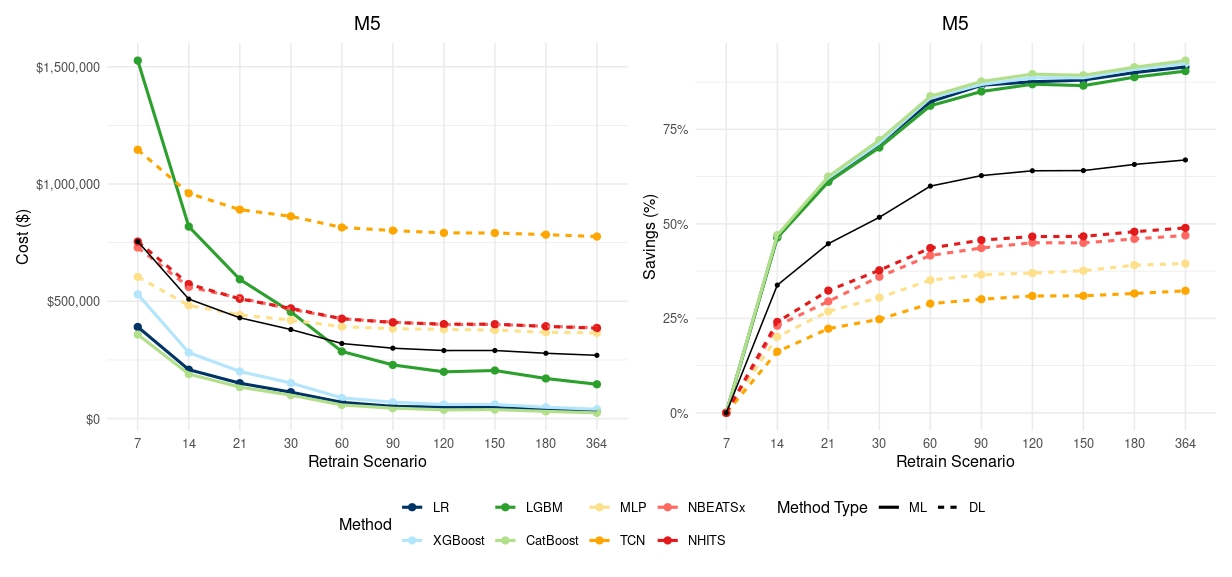}
    \caption{Daily data estimated costs and percentage savings for each method and 
    retrain scenario combination. The black line represents the average. Costs
    are in real values, while savings are expressed in relative terms 
    with respect to the baseline scenario, \(r = 7\).}
    \label{fig:m5_cost_plot}
\end{figure}

\begin{figure}
    \centering
    \includegraphics[scale=0.55]{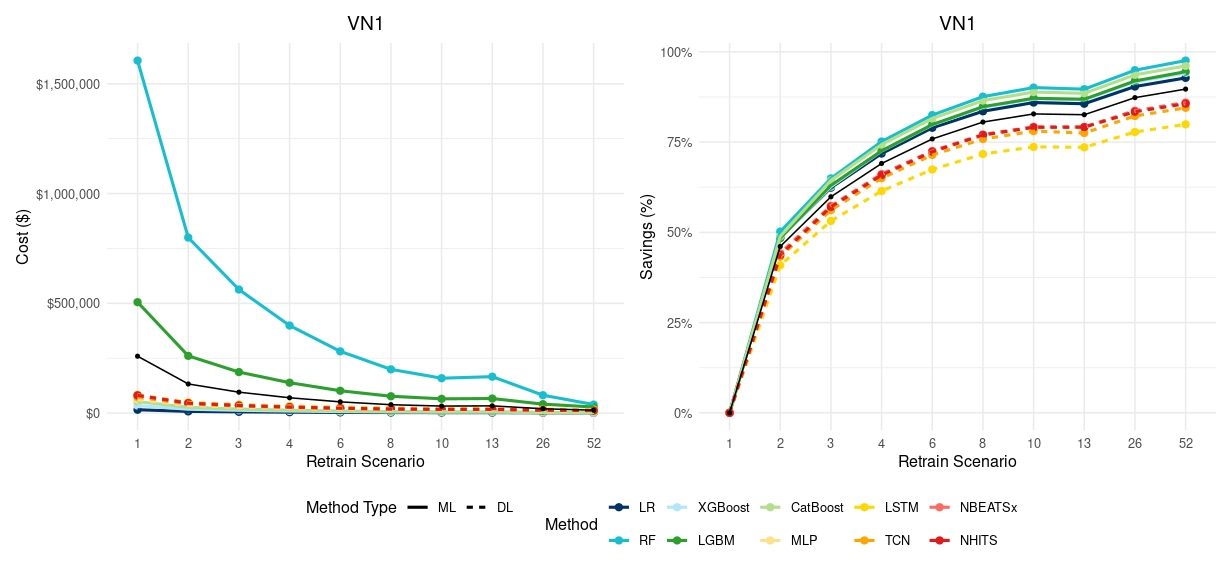}
    \caption{Weekly data estimated costs and percentage savings for each method and 
    retrain scenario combination. The black line represents the average. Costs
    are in real values, while savings are expressed in relative terms 
    with respect to the baseline scenario, \(r = 1\).}
    \label{fig:vn1_cost_plot}
\end{figure}

Overall, the results of Figures \ref{fig:rmsse_plot}, \ref{fig:mql_plot}, and
\ref{fig:ct_plot}, combined with the Friedman-Nemenyi tests for statistical 
significance (see Supplementary material), suggest that 
retraining the global models less frequently does not harm (or even improves) 
forecast accuracy, while at the same time significantly reducing the computing 
time of producing the forecasts. Thus, extending the retraining time from 
the standard practice of continuous retaining to some level of periodic retraining 
allows for effective management of the computing time, and in turn, the costs of forecasting.
Indeed, as already observed, the computational time can be directly translated 
into actual costs for the company. Following \citet{forsubopt} and
\citet{occamrazor}, we assumed some standard costs for computing services 
to estimate the costs of forecasting associated with each retrain scenario. 
Note that costs and savings are normalized by the number of series considered 
in each dataset so that they can be directly compared and 
conclusions can be drawn in terms of the frequency of the time series.
Figures \ref{fig:m5_cost_plot}, and \ref{fig:vn1_cost_plot} 
show the costs and associated savings for a large retailer, 
given a fixed computing service cost of \$3.5/hour, 200,000 unique SKUs 
and 5,000 stores, which is approximately the size of the forecasting problem 
of Walmart \citep{tsupdate1}.
As expected, the costs decrease exponentially with less frequent retraining.
For daily data, forecasting with machine learning models usually costs less than 
with deep learning models.  
On average, the continuous retraining scenario costs approximately \$750,000 and 
this cost drops down to almost \$250,000 in the no retraining scenario, implying
direct savings of more than 60\%.
Moreover, machine learning models allow for higher savings. Indeed, 
even if the models have to be updated, going from retraining every week to retraining 
every month allows to get direct savings of almost 75\%, while it can be less than 
30\% for deep learning models. This implies that machine learning methods
are a marginally better choice to reduce the costs of forecasting when coupled with
less frequent model retraining as the frequency of the data increases.
For weekly data, these differences are less evident, meaning that machine learning
and deep learning models usually have very similar costs and savings profiles 
(without considering the Random Forest model, which is usually more than 10 times
slower than other methods). This is to be expected since lower frequencies also imply
smaller datasets and lower computing time. In this case, the average cost 
of forecasting under a continuous retraining scenario is \$250,000 (1/3 of that of
daily data), and it drops to \$15,000 in the least frequent retraining 
scenario, producing direct savings of 90\%. Nevertheless, for weekly data too,
moving from retraining a model every week (\(r = 1\)) to just once a month 
(\(r = 4\)) allows to reduce the cost of almost 75\%.

It may be argued that for a large retailer (like Walmart), these costs (and savings) 
may be negligible. However, it has to be noted that the cost reduction obtained 
from less frequent retraining comes with no shortcomings in terms of forecasting 
accuracy (especially for point forecasting). Indeed, as already mentioned, models 
under a periodic retraining scenario result in at least the same (if not even better) 
forecasting performance compared to the usual practice of continuous retraining. 
Moreover, periodic retraining may be a good practice even when probabilistic forecasts 
are needed, balancing costs and accuracy effectively.

\begin{figure}
    \centering
    \includegraphics[scale=0.55]{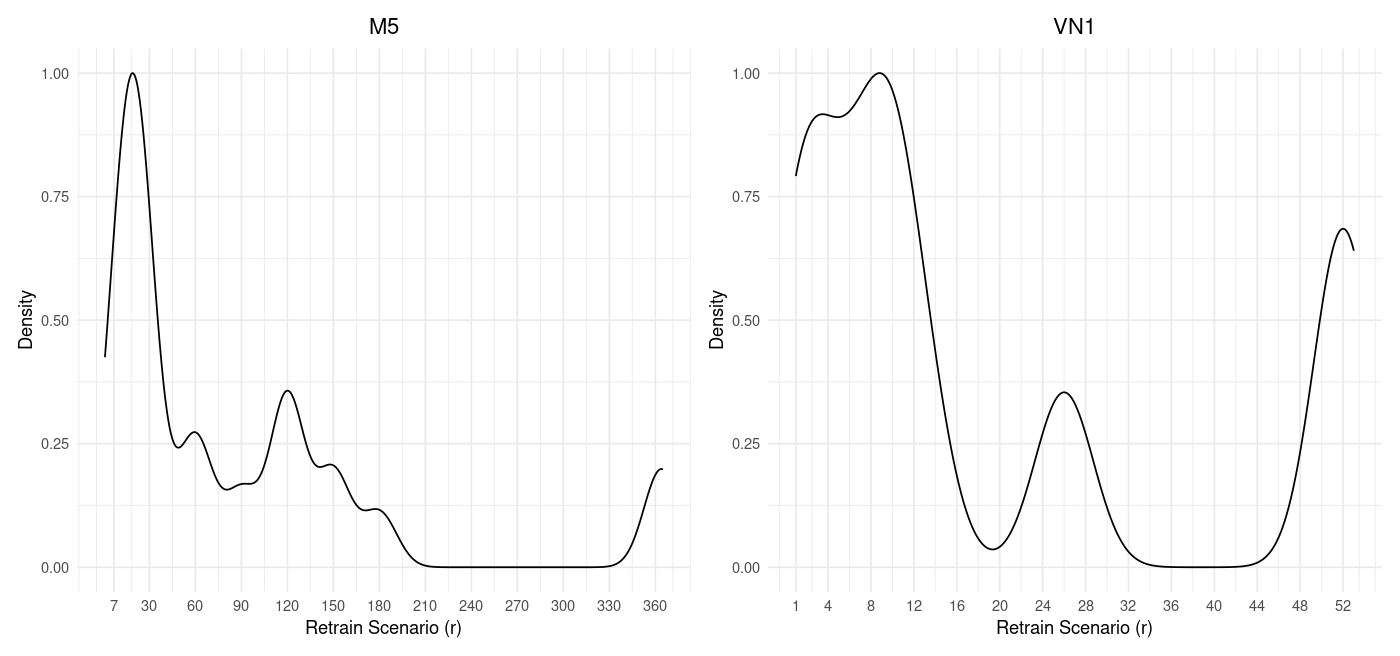}
    \caption{Overall optimal retraining frequency based on RMSSE. For each series, 
    the optimal frequency is identified based on the average accuracy reported across
    the evaluation rounds per update frequency, considering all the models.}
    \label{fig:optimal_rmsse}
\end{figure}

Finally, in this study, since we are dealing with global models, we assumed 
a fixed model retraining frequency across all forecast origins and time series. 
While our results showed no significant difference in forecasting accuracy 
when reducing the retraining frequency, we believe there is potential to 
further improve accuracy, especially when adopting some form of periodic 
retraining, for instance, by allowing the retraining frequency to vary dynamically. 
Such adaptive strategies could involve automated procedures that monitor 
forecast error levels and trigger updates only when errors exceed a certain 
threshold. This approach would still yield computational savings, albeit smaller, 
since the model re-training would not be required at every review point.
In this direction, Figure \ref{fig:optimal_rmsse} illustrates the "optimal" 
retraining frequency identified for each series in the M5 and VN1 datasets, 
considering all the models.
The results confirm that although certain series may benefit from frequent 
retraining, in most cases, a less frequent retraining of the models leads to 
better accuracy and efficiency. Indeed, for point forecasting, the optimal 
retraining frequency was found to be between 3 and 4 weeks (approximately 
21 observations) for the M5 dataset, and around 8 to 10 weeks for the VN1 
dataset. In the context of probabilistic forecasting, instead, the optimal 
retraining intervals were shorter, around 2 weeks for M5 and 4 weeks for VN1 
(see Supplementary material), suggesting that uncertainty estimation may
require more frequent updates to maintain accurate levels. 
Moreover, while these results do not differentiate among the global 
forecasting methods used in the experiment, individual model differences exist, 
implying that even further improvements may be obtained by relying on 
specific model (see Supplementary material).
These findings suggest that exploring intermediate retraining strategies and 
determining the optimal retraining frequency offers promising avenues for 
reducing the computational cost of the forecasting process while maintaining 
a high level of accuracy.

\section{Conclusions} \label{sec:conc}

In this study, we went beyond the traditional evaluation of forecasting models 
by also exploring the computational cost associated with generating forecasts.
We examined the effects of retraining frequency on the accuracy 
and computational efficiency of global time series forecasting models. By 
systematically evaluating various retraining scenarios, ranging from continuous 
retraining to no retraining at all, across multiple machine learning and deep 
learning models, we aimed to address a key question in forecasting: is frequent 
retraining necessary for maintaining the high predictive performance of global models? 
Our findings, derived from an extensive evaluation of ten different global 
models across two large-scale, real-world retail datasets, challenge the 
conventional wisdom of continuous model retraining. Indeed, our results indicate 
that the common practice of continuous retraining may not always be 
justified, and that periodic retraining strategies can offer substantial benefits 
in terms of computational efficiency without a significant loss in accuracy.

Our empirical evaluation provides evidence that the forecasting accuracy of 
global models remains stable even when the retraining frequency is significantly 
reduced. 
For point forecasting, the root mean squared scaled error (RMSSE) results 
indicate that models trained less frequently do not exhibit substantial performance 
degradation. In fact, in some cases, particularly for periodic retraining frequencies, 
we observe marginal improvements in accuracy.
For probabilistic forecasting, the results are slightly more nuanced. While less 
frequent retraining does lead to minor reductions in accuracy as measured by the scaled
Multi-Quantile Loss (SMQL), the degradation is relatively small (typically within 
5-6\%). This implies that for most practical applications, especially those where 
forecasting costs are key considerations, periodic retraining provides a favorable 
balance between accuracy and efficiency.
This suggests that global models, that learn shared dynamics across multiple time 
series, are robust to the absence of frequent updates, especially in contexts with 
relatively stable demand patterns.

Considering periodic retraining scenarios allows for effective management of the 
computing time, and in turn, the costs of forecasting. Indeed, we have shown how the 
computational time can be directly translated into actual costs for the company.
One of the most significant findings of this study is the exponential reduction in 
computational costs as retraining frequency decreases. The computational time (CT) 
analysis demonstrates that moving from continuous retraining to a monthly retraining 
schedule can reduce computational costs by approximately 75\%. In the extreme case of 
no retraining, the cost reductions approach 90\%, representing a resource-saving 
opportunity with no shortcomings.
Furthermore, the cost analysis performed on realistic retail settings (e.g., 
Walmart-sized operations) reveals that periodic retraining can yield relevant 
financial savings. The estimated costs of forecasting, which scale with computing 
service expenses, decrease sharply when retraining is performed less frequently. 
This suggests that organizations can achieve economic benefits by optimizing their 
retraining schedules without compromising the forecast quality.

The implications of these findings are highly relevant for both academic research 
and industry applications. First, they challenge the prevailing assumption that 
forecasting models require frequent updates to maintain high predictive performance. 
Instead, our results suggest that global models remain effective over extended periods, 
meaning that retraining can be strategically planned rather than performed continuously.
For practitioners, our results provide concrete guidelines on how often retraining 
should occur. In general, retraining every month appears to be a viable option to balance 
probabilistic accuracy and costs. Instead, if the forecasting objective is on point 
forecast, then even longer retraining scenarios may be adopted. 
Moreover, our results shed some light on the computational comparison between machine 
learning and deep learning models. We found that the former benefits more from 
less frequent retraining as the frequency of the data increases. This implies that, 
for large-scale applications, like the retail industry, where the forecasts of many 
different SKUs have to be provided regularly, machine learning models are a marginally 
better choice to reduce the costs of forecasting when coupled with less frequent model 
retraining strategies as the frequency of the data increases.
These insights also have broader implications for sustainability in machine learning 
and AI-driven forecasting. The computational cost savings associated with reduced 
retraining directly translate into lower energy consumption, making forecasting 
operations more environmentally sustainable. This aligns with recent discussions on 
"Green AI" which emphasizes the importance of optimizing computational resources to 
reduce the environmental impact of machine learning applications.

While our findings provide strong evidence for the feasibility of less frequent 
retraining in global forecasting models, some limitations remain. First, our study 
focused on two large retail datasets (M5 and VN1), which may not fully capture all 
possible time series patterns. Moreover, in this paper, we assumed that the data-generating
process remains stable without exhibiting significant trends or concept 
drifts, but in some real-world applications, this assumption may simply not be valid.
Future research could extend this analysis to other domains, such as financial time 
series or industrial production data, to verify the general validity of these findings.
Furthermore, our study did not explore adaptive retraining strategies, where models are 
updated only when a significant drift in data distribution is detected. Such adaptive 
approaches could provide an even more refined balance between accuracy and computational 
efficiency. Future work could investigate the integration of drift detection mechanisms 
to optimize retraining schedules dynamically.
Another direction for future studies may be the analysis of the effect of different 
retraining scenarios on the stability of global models, that is, answering the question 
\textit{"Do global models produce stable forecasts even if not continuously retrained?"}.
Extending the analysis to some very recent global forecasting methods that demonstrate
state-of-the-art performance (e.g., transformers) would also be a natural path for future
research.
Finally, we encourage further studies to explore the interaction between retraining 
frequency and hyperparameter optimization. Indeed, different models may respond 
differently to retraining strategies depending on their hyperparameter configurations, 
and understanding these interactions could lead to even more efficient forecasting 
strategies.

In summary, our study shows that frequent retraining is not necessarily required 
to maintain high forecasting accuracy in global models. Less frequent retraining can 
significantly reduce computational costs while maintaining competitive forecasting 
performance. These findings have important implications for organizations seeking to 
optimize their forecasting pipelines, offering a pathway toward more efficient and 
sustainable forecasting practices. By shifting from a continuous retraining paradigm to 
a periodic retraining approach, businesses can achieve relevant cost savings while still
ensuring high predictive performance.





\bibliography{main}

\end{document}